\documentclass[journal,10pt,draftclsnofoot,onecolumn]{IEEEtran}
\usepackage{color}
\usepackage[numbers,sort&compress]{natbib}

\ifCLASSINFOpdf
\else
\fi
\usepackage[cmex10]{amsmath}

\usepackage{graphicx,epic,epsfig,amsmath,latexsym,amssymb,mathrsfs,verbatim,subfigure,color}

\newtheorem{definition}{Definition}

\newtheorem{theorem}[definition]{Theorem}
\newtheorem{corollary}[definition]{Corollary}

\newtheorem{example}[definition]{Example}

\newcommand{\iden}{1 \hspace{-1.0mm}  {\bf l}}
\newcommand{\bra}[1]{\langle#1|}
\newcommand{\ket}[1]{|#1\rangle}

\def\squareforqed{\hbox{\rlap{$\sqcap$}$\sqcup$}}
\def\qed{\ifmmode\squareforqed\else{\unskip\nobreak\hfil
\penalty50\hskip1em\null\nobreak\hfil\squareforqed
\parfillskip=0pt\finalhyphendemerits=0\endgraf}\fi}
\def\endenv{\ifmmode\;\else{\unskip\nobreak\hfil
\penalty50\hskip1em\null\nobreak\hfil\;
\parfillskip=0pt\finalhyphendemerits=0\endgraf}\fi}

\long\def\ignore#1{}

\newcommand{\nc}{\newcommand}
\nc{\rnc}{\renewcommand}
\nc{\beq}{\begin{equation}}
\nc{\eeq}{{\end{equation}}}
\nc{\beqa}{\begin{eqnarray}}
\nc{\eeqa}{\end{eqnarray}}
\nc{\lbar}[1]{\overline{#1}}
\nc{\proj}[1]{| #1\rangle\!\langle #1 |}
\nc{\avg}[1]{\langle#1\rangle}
\nc{\Rank}{\operatorname{Rank}}
\nc{\smfrac}[2]{\mbox{$\frac{#1}{#2}$}} \nc{\Tr}{\operatorname{Tr}}
\nc{\tr}{\operatorname{Tr}} \nc{\id}{\operatorname{id}}
\nc{\1}{\openone} \nc{\ox}{\otimes} \nc{\dg}{\dagger}
\nc{\dn}{\downarrow} \nc{\cA}{{\cal A}} \nc{\cB}{{\cal B}}
\nc{\cC}{{\cal C}} \nc{\cD}{{\cal D}} \nc{\cE}{{\cal E}}
\nc{\cF}{{\cal F}} \nc{\cG}{{\cal G}} \nc{\cH}{{\cal H}}
\nc{\cI}{{\cal I}} \nc{\cJ}{{\cal J}} \nc{\cK}{{\cal K}}
\nc{\cL}{{\cal L}} \nc{\cM}{{\cal M}} \nc{\cN}{{\cal N}}
\nc{\cO}{{\cal O}} \nc{\cP}{{\cal P}} \nc{\cQ}{{\cal Q}}
\nc{\cR}{{\cal R}} \nc{\cS}{{\cal S}} \nc{\cT}{{\cal T}}
\nc{\cX}{{\cal X}} \nc{\cY}{{\cal Y}} \nc{\cZ}{{\cal Z}}
\nc{\supp}{{\operatorname{supp}}} \nc{\var}{\operatorname{var}}
\nc{\rar}{\rightarrow} \nc{\lrar}{\longrightarrow}
\nc{\polylog}{\operatorname{polylog}}

\begin{document}
%
\title{Quantum Error-Correcting Codes\\for Qudit Amplitude Damping}

\author{Markus Grassl, Linghang Kong, Zhaohui Wei, Zhang-Qi Yin, Bei
  Zeng
\thanks{Part of this work was presented at the 2014 IEEE Symposium on
  Information Information Theory.}
\thanks{Markus Grassl is with Universit\"at Erlangen-N\"urnberg and
  the Max Planck Institute for the Science of Light, 91058 Erlangen, Germany}
\thanks{Zhaohui Wei is with the Center for Quantum Technologies, National University of Singapore, 117543, Singapore and School of Physical and
Mathematical Sciences, Nanyang Technological University, 639798,
Singapore}
\thanks{Linghang Kong and Zhang-Qi Yin are with the Center for Quantum Information, Institute for Interdisciplinary Information Sciences,
Tsinghua University, Beijing 100084, P. R. China}
\thanks{Bei Zeng is with the Department of Mathematics $\&$ Statistics, University of Guelph, Guelph, ON, N1G 2W1, Canada
and the Institute for Quantum Computing, University of Waterloo, Waterloo,  Ontario, N2L 3G1, Canada}}


%



\makeatletter
\def\@IEEEaftertitletext{\vspace*{-5mm}}
\makeatother
\abovedisplayskip0.75\abovedisplayskip
\belowdisplayskip0.75\belowdisplayskip
\intextsep0.5\intextsep
\floatsep0.5\floatsep
\abovecaptionskip0.5\abovecaptionskip

\maketitle



\begin{abstract}
Traditional quantum error-correcting codes are designed for the
depolarizing channel modeled by generalized Pauli errors occurring
with equal probability.  Amplitude damping channels model, in
general, the decay process of a multilevel atom or energy
dissipation of a bosonic system at zero temperature.  We discuss
quantum error-correcting codes adapted to amplitude damping channels
for higher dimensional systems (qudits).  For multi-level atoms, we
consider a natural kind of decay process, and for bosonic systems,
we consider the qudit amplitude damping channel obtained by
truncating the Fock basis of the bosonic modes to a certain maximum
occupation number.  We construct families of single-error-correcting
quantum codes that can be used for both cases.  Our codes have
larger code dimensions than the previously known
single-error-correcting codes of the same lengths.  Additionally, we
present families of multi-error correcting codes for these two
channels, as well as generalizations of our construction
technique to error-correcting codes for the qutrit $V$ and $\Lambda$
channels.
\end{abstract}


\section{Introduction}
\label{sec:intro}

For a $q$-level quantum system with Hilbert space $\mathbb{C}^q$,
called a qudit, the most general physical operations (or quantum
channels) allowed by quantum mechanics are completely positive, trace
preserving linear maps which can be represented in the following Kraus
decomposition form $\mathcal{N}(\rho)=\sum_k E_k \rho E_k^\dag$, where
the matrices $E_k$ are called Kraus operators of the quantum channel
$\mathcal{N}$ satisfying the trace-preserving condition $\sum_k
E_k^{\dag} E_k=\iden$.

In designing error-correcting codes for protecting messages carried
by $n$ qudits sent through a channel $\cN$, it is usually assumed
that the errors to be corrected are completely random, with no
knowledge other than that they affect different qudits independently
\cite{nielsenchuang,thesis:gottesman}. The corresponding channel
$\mathcal{N}$ is the depolarizing channel which can be modeled by
uniformly distributed error-operators given by generalized Pauli
operators \cite{GKP01,PZ88,Kni96} $(X_q)^a(Z_q)^b$, for
$a,b\in\{0,1,\ldots,q-1\}$, where $X_q\ket{s}=\ket{s+1\mod q}$,
$Z_q\ket{s}=\omega^s\ket{s}$, and $\omega=\exp({2\pi i}/{q})$. When
it is clear from the context, we may just write $X$ and $Z$,
dropping the index $q$.

However, if further information about the error process is
available, more efficient codes can be designed.  Indeed, in many
physical systems, the noise is likely to be unbalanced between
amplitude ($X$-type) errors and phase ($Z$-type) errors. Recently a
lot of attention has been put into designing codes for this
situation and into studying their fault tolerance properties
\cite{aliferis-biased-2007,evans-2007,fletcher-ad,Ioffe-bias,Martin:ISIT2008,WFL+10}.
All these results use error models described by Kraus operators that
are generalized Pauli operators, but for those error models, the $X$-type
errors (i.e., non-diagonal Pauli matrices) happen with probability
$p_x$ which might be different from the probability $p_z$ that
$Z$-type errors (i.e., diagonal Pauli matrices) happen.  The quantum
channels described by this kind of noise are called asymmetric
channels.

A closer look at the real physical process of amplitude damping noise
shows that one needs to go even further, beyond Kraus operators of
Pauli type.  To be more precise, for $q=2$, the qubit amplitude
damping (AD) channel is given by the Kraus operators \cite{Chuang1997}
\begin{equation}
A_0=\begin{pmatrix} 1 & 0 \\0 &
\sqrt{1-\gamma} \end{pmatrix}
\quad\text{and}\quad A_1=\begin{pmatrix} 0 &
\sqrt{\gamma} \\0 & 0 \end{pmatrix}.
\label{eq:ADKraus}
\end{equation}
Since the error model of the qubit AD channel is not described by
Pauli-type Kraus operators, the known techniques dealing with Pauli
errors result in codes with non-optimal parameters.  Several new
techniques for the construction of codes adapted to this type of noise
with non-Pauli Kraus operators, and the qubit AD channel in
particular, have been developed
\cite{Chuang1997,fletcher-ad,LS07,LNCY97,SSS+11}.  Systematic methods
for constructing high performance single-error-correcting
codes~\cite{LS07,SSS+11} and multi-error-correcting
codes~\cite{DGJZ10} have been found.

In this paper, we discuss constructions of quantum codes for AD
channels of general qudit systems. Unlike the qubit case, where the AD
channel is unique, for qudit systems there are different AD channels
associated with different physical systems.
We will focus on two different models: multi-level atoms with a
natural kind of decay process, and bosonic systems obtained by
truncating the Fock basis of the bosonic modes to the maximum
occupation number $q-1$ for a single bosonic mode.

\section{The Amplitude Damping Channel}

For two-level atoms, the decay process at zero temperature is
described by the Kraus operators $A_0,A_1$ as given in
Eq. \eqref{eq:ADKraus}. For multi-level atoms, there are different
kinds of decay processes at zero temperature. One natural decay
process is the cascade structure $\Xi$, where the decay process is
governed by the master equation~\cite{SZ97,Puri01}
\begin{alignat}{5}
\frac{d\rho}{d\tau}&=
&\sum_{\rlap{$\scriptstyle1\leq i\leq q-1$}}
k_i\left(2\sigma_{i-1,i}^{-}\rho\sigma_{i-1,i}^{+}
-\sigma_{i-1,i}^{+}\sigma_{i-1,i}^{-}\rho
-\rho\sigma_{i-1,i}^{+}\sigma_{i-1,i}^{-}\right).
\end{alignat}
Here $\{\ket{i}\}_{i=0}^{q-1}$ is a basis of the Hilbert space
$\mathbb{C}^q$, $\sigma_{i-1,i}^{-}=|i-1\rangle\langle i|$ and
$\sigma_{i-1,i}^{+}=|i\rangle\langle i-1|$.

The solution to this master equation gives the Kraus expression
\begin{equation}
\Xi(\rho)
=A_0\rho A_0^{\dag}+\sum_{0\leq i<j\leq q-1}A_{ij}\rho A_{ij}^{\dag},
\end{equation}
where $A_{ij}=\sqrt{\gamma_{ij}}|i\rangle\langle j|$ with positive
coefficients $\gamma_{ij}$, and $A_0$ is a diagonal matrix with its
elements determined by $A_0^{\dag}A_0+\sum_{0\leq i<j\leq
  q-1}A_{ij}^{\dag}A_{ij}=I$. Furthermore, when the decay time $\tau$ is
small, $\gamma_{ij}$ is of order $\tau^\ell$ for any $j=i+\ell$,
$\ell>0$. As a consequence, $A_0$ is of order $\tau$, and $A_{ij}$ is of
order $\tau^{\ell/2}$ for any $j=i+\ell$, $\ell>0$. This is intuitively
sound as for the cascade structure, the first order transition always
happens from $\ket{i+1}$ to $\ket{i}$.

As an example, for three-level atoms, i.e., $q=3$, we have
\begin{alignat*}{5}
A_{01}&=\sqrt{\gamma_{01}}|0\rangle\langle1|,\\
A_{02}&=\sqrt{\gamma_{02}}|0\rangle\langle2|,
&&A_{12}=\sqrt{\gamma_{12}}|1\rangle\langle2|,\\
A_0&=|0\rangle\langle0|+\sqrt{1-\gamma_{01}}&&|1\rangle\langle1|
+\sqrt{1-\gamma_{02}-\gamma_{12}}|2\rangle\langle2|,
\end{alignat*}
where
\begin{alignat*}{5}
\gamma_{01}&=2k_2\tau+O(\tau^2),\\
\gamma_{02}&=2 k_1 k_2 \tau^2+O(\tau^3),\\
\gamma_{12}&=2k_1\tau+O(\tau^2),
\end{alignat*}
for $k_1\neq k_2$. The values of $\gamma_{ij}$ are slightly different
for $k_1=k_2$, but the order of $\gamma_{ij}$ in terms of $\tau$ remains
the same.

The channel $\mathcal{A}$ describing energy dissipation of a bosonic
system at zero temperature is discussed in \cite{Chuang1997}.  The
Kraus operators are given by
\begin{equation}
\label{eq:Ak}
A_k=\sum_{r=k}^{q-1}\sqrt{\binom{r}{k}}\sqrt{(1-\gamma)^{r-k}\gamma^k}\ket{r-k}\bra{r},
\end{equation}
where $q-1$ is the maximum occupation number of a single bosonic mode,
and $k=0,1,\ldots,q-1$. The parameter $\gamma$ is of first order in
terms of the decay time $\tau$, i.e., $\gamma=c\tau+O(\tau^2)$. As a
consequence, the non-identity part of $A_0$ is of order $\tau$, and $A_k$
is of order $\tau^{k/2}$ for $1\leq k\leq d-1$.

For instance, for the qubit case, i.e., $q=2$, we have the qubit
amplitude channel given by Eq. \eqref{eq:ADKraus}.  For $q=3$, we have
\begin{alignat*}{5}
A_0&=\ket{0}\bra{0}+\sqrt{1-\gamma}\ket{1}\bra{1}+(1-\gamma)\ket{2}\bra{2},\\
A_1&=\sqrt{\gamma}\ket{0}\bra{1}+\sqrt{2\gamma(1-\gamma)}\ket{1}\bra{2},\\
\text{and}\quad A_2&={\gamma}\ket{0}\bra{2}.
\end{alignat*}

Note that for $q=3$, the non-diagonal Kraus operators of the channel
$\mathcal{A}$ for bosonic systems are linear combinations of the Kraus
operators of the channel $\Xi$. Hence codes correcting errors of the
channel $\Xi$ are also codes for the channel $\mathcal{A}$.

\section{Error Correction Criteria}

A quantum error-correcting code $Q$ is a subspace of
$(\mathbb{C}^q)^{\otimes n}$, the space of $n$ qudits.  For a
$K$-dimensional code space spanned by the orthonormal basis
$\ket{\psi_i}$, $i=1,\ldots,K$, and a set of errors $\mathcal{E}$,
there is a physical operation correcting all elements $E_k \in
\mathcal{E}$ if the error correction conditions
\cite{bennett-1996-54,KL97} are satisfied:
\begin{equation}\label{eq:KnillLaflamme}
\forall i,j,k,l\colon
\bra{\psi_i} E_k^\dag E_l \ket{\psi_j} =\delta_{ij}\alpha_{kl},
\end{equation}
where $\alpha_{kl}$ depends only on $k$ and $l$.
A code is said to be pure with respect to some set of errors
$\mathcal{E}$ if $\alpha_{kl}=0$ for $k\ne l$.
A $K$-dimensional code with length $n$ is denoted by $(\!(n,K)\!)$.

For the AD channels $\Xi$ and $\mathcal{A}$, if the decay time $\tau$
is small, we would like to correct the leading order errors that occur
during amplitude damping \cite{BO10,Beny11}.  Similar as for the qubit
case \cite[Section 8.7]{thesis:gottesman}, we will show below that in
order to improve the fidelity of the transmission through the AD
channel $\Xi$ or $\mathcal{A}$ from $1-O(\tau)$ to $1-O(\tau^{t+1})$,
i.e., to correct $t$ errors, it is sufficient to be able to detect $t$
errors of type $A_0$ and to correct up to $t$ errors of type $A_{ij}$
with $j>i$ of total order $\tau^{t/2}$ for the channel $\Xi$ (or to
correct up to $t$ errors of type $A_i$ with $i>0$ for the channel
$\mathcal{A}$).  We will then say that such a code corrects $t$
amplitude damping errors since it improves the fidelity, just as much
as a true $t$-error-correcting code would for the same channel.  This
is a direct consequence of the following sufficient condition for
approximate error correction.

\begin{theorem}
\label{th:approx}\label{thm:approx}
Assume we are given a quantum channel with Kraus operators $E_k$ that
have a series expansion in terms of $\sqrt{\tau}$ for a parameter
$\tau$.  A quantum code $\cQ$ with orthonormal basis $\{\ket{c_i} :
i = 1, \ldots , K\}$ corrects errors up to order $O(\tau^t)$ if the
following conditions are fulfilled for all basis states $\ket{c_i},
\ket{c_j}$ and all error operators $E_k, E_l$:
\begin{equation}\label{eq:approx_condition}
\bra{c_i}E_k^\dag E_l\ket{c_j} = \delta_{ij}\nu_{kl} + O(\tau^{t+1})
\end{equation}
\end{theorem}

\begin{IEEEproof}
Assume we are given a quantum channel with Kraus operators
$E_k(\tau)$ that depend on some small parameter $\tau$. We expand
the operators in terms of $\sqrt{\tau}$ as
\begin{equation}
E_k(\tau)=\sum_{m\ge 0} E_{k m}\tau^{m/2} \label{eq:serial}.
\end{equation}
This leads to the following description of the channel:
\begin{alignat}{5}
\rho\mapsto&
\sum_{k} E_k\rho E_k^\dagger
&=\sum_{k}\sum_{m\ge 0}\sum_{\mu\ge 0} E_{km} \rho E_{k\mu}^\dagger\tau^{(m+\mu)/2}.
\end{alignat}

Given a quantum code $\cQ$ with orthonormal basis $\{\ket{c_i}\colon
i=1,\ldots,K\}$, the conditions for perfect error correction are
\begin{equation}
\bra{c_i}E_k^\dagger(\tau)E_l(\tau)\ket{c_j}=\delta_{ij}\alpha_{k  l}.
\end{equation}
Using the expansion of the Kraus operators in terms of $\sqrt{\tau}$, we
get
\begin{equation}
\sum_{m,\mu\ge 0}\bra{c_i}E_{km}^\dagger E_{\ell\mu}\ket{c_j}\tau^{(m+\mu)/2}
=\delta_{ij}\alpha_{k\ell}.
\end{equation}

We are looking for sufficient conditions such that the
residual error with respect to $\rho$ is of order $O(\tau^{t+1})$ for some $t>0$.

We  write each Kraus operator $E_k=B_k+C_k+D_k$ as sum of three terms, where
\begin{alignat}{5}
B_k&=\sum_{m=0}^{t} E_{km}\tau^{m/2},\label{eq:def_B_k}\\
C_k&=\sum_{m=t+1}^{2t} E_{km}\tau^{m/2},\label{eq:def_C_k}\\
D_k&=\sum_{m>2t} E_{km}\tau^{m/2}.\label{eq:def_D_k}
\end{alignat}
Then the original channel $\mathcal{N}$ can be written as
\begin{alignat}{5}
\mathcal{N}(\rho)=&\sum_k (B_k+C_k+D_k)\rho(B_k+C_k+D_k)^\dagger\\
=&\sum_k B_k\rho B_k^\dagger\label{noise_B_k}\\
&+\sum_k B_k\rho C_k^\dagger+ C_k\rho B_k^\dagger \label{noise_B_C}\\
&+\sum_k B_k\rho D_k^\dagger+D_k\rho B_k^\dagger \label{noise_B_D}\\
&+\sum_k (C_k+D_k)\rho(C_k+D_k)^\dagger.\label{noise_C_D}
\end{alignat}
Condition \eqref{eq:approx_condition} implies that
\begin{alignat}{5}
\bra{c_i}B_k^\dagger B_l\ket{c_j}&{}=\delta_{ij}\lambda_{kl},\label{eq:correction}\\
\bra{c_i}B_k^\dagger C_l\ket{c_j}&{}=\delta_{ij}\mu_{kl}+O(\tau^{t+1}).\label{eq:detection}
\end{alignat}
In particular, the error operators $B_k$ can be perfectly corrected.
We first define the projection operator onto one of the
spaces\footnote{We may, without loss of generality, use linear
  combinations of the original error operators $B_k$ such that these
  spaces become mutually orthogonal.}  spanned by
$\{B_k\ket{c_i}\colon i=1,\ldots,K\}$:
\begin{alignat}{5}
P_{B_k}&=\sum_i B_k\ket{c_i}\bra{c_i}B_k^\dagger,
\end{alignat}
and the partial isometry that maps
$B_k\ket{c_i}$ to $\ket{c_i}$:
\begin{alignat}{5}
U_{B_k}&=\sum_i \ket{c_i}\bra{c_i}B_k^\dagger.
\end{alignat}
We compute
\begin{alignat}{5}
U_{B_k}P_{B_k}&=\left(\sum_i \ket{c_i}\bra{c_i}B_k^\dagger\right)
\left(\sum_j B_k\ket{c_j}\bra{c_j}B_k^\dagger\right)\\
&=\sum_{i,j} \ket{c_i}\bra{c_i}B_k^\dagger B_k\ket{c_j}\bra{c_j}B_k^\dagger\\
&=\lambda_{kk}\sum_i \ket{c_i}\bra{c_i}B_k^\dagger.
\end{alignat}
The last step follows from the fact that the error operators $B_k$ can
be perfectly corrected, which also determines the constant
$\lambda_{kk}$.  Then the partial correction operator
$\mathcal{R}_{\text{part}}$ is given by
\begin{alignat}{5}
\mathcal{R}_{\text{part}}(\rho)&=\sum_k |\lambda_{kk}|^2
\sum_{i,j}\ket{c_i}\bra{c_i}B_k^\dagger\,\rho\, B_k\ket{c_j}\bra{c_j}.
\end{alignat}
For a general state
\begin{alignat}{5}
\rho_Q=\sum_{i,j}\alpha_{ij}\ket{c_i}\bra{c_j}
\end{alignat}
in the quantum code $Q$, the term \eqref{noise_B_k} of
$\mathcal{N}(\rho_Q)$ reads
\begin{alignat}{5}
\mathcal{N}_{B_k}(\rho_Q)&=\sum_k\sum_{i,j}\alpha_{ij}B_k\ket{c_i}\bra{c_j}B_k^\dagger.
\end{alignat}
Since the error operators $B_k$ can be perfectly corrected (implied by
Eq.~\eqref{eq:correction}), it can be shown that applying the partial
recovery operator to $\mathcal{N}_{B_k}(\rho_Q)$ yields a state
$\lambda\rho_Q$ that is proportional to the original state $\rho_Q$.
Hence after partial recovery we have
\begin{alignat}{5}
\mathcal{R}_{\text{part}}(\mathcal{N}(\rho_Q))&=\lambda\rho_Q+\mathcal{S}(\rho_Q),
\end{alignat}
where the map $\mathcal{S}$ is given by the application of the
partial recovery operator to the terms given in \eqref{noise_B_C}, \eqref{noise_B_D} and \eqref{noise_C_D}.
The summands \eqref{noise_B_D} and \eqref{noise_C_D} are all of order
$O(\tau^{t+1})$, so we can ignore them, but \eqref{noise_B_C}
contains terms of order $\tau^{l/2}$ for $t<l\le 2t$.  Applying the
partial recovery operator to \eqref{noise_B_C} and using
\eqref{eq:detection} results in the state 

\begin{alignat*}{5}
\mathcal{R}_{\text{part}}&\left(\sum_k B_k\rho_Q C_k^\dagger+ C_k\rho_Q B_k^\dagger\right)\\
&=\sum_k |\lambda_{kk}|^2\sum_{i,j}\ket{c_i}\bra{c_i}B_k^\dagger\left(\sum_l B_l\rho_Q
C_l^\dagger+ C_l\rho_Q B_l^\dagger\right) B_k\ket{c_j}\bra{c_j}\\
&=\sum_{k,l} |\lambda_{kk}|^2\sum_{i,j}\sum_{i',j'}\alpha_{i',j'}\Bigl(
\ket{c_i}\underbrace{\bra{c_i}B_k^\dagger B_l\ket{c_{i'}}}_{\lambda_{kl}\delta_{ii'}}\bra{c_{j'}} C_l^\dagger B_k\ket{c_j}\bra{c_j}
+\ket{c_i}\bra{c_i}B_k^\dagger C_l\ket{c_{i'}}\underbrace{\bra{c_{j'}} B_l^\dagger B_k\ket{c_j}}_{\lambda_{lk}\delta_{jj'}}\bra{c_j}
\Bigr)\\
&=\sum_{k,l} |\lambda_{kk}|^2\sum_{i',j'}\alpha_{i',j'}\Bigl(
\sum_j\lambda_{kl}\underbrace{\bra{c_{j'}} C_l^\dagger B_k\ket{c_j}}_{\delta_{jj'}\mu_{kl}^*+O(\tau^{t+1})}\ket{c_{i'}}\bra{c_j}
+\sum_i\lambda_{lk}\underbrace{\bra{c_i}B_k^\dagger C_l\ket{c_{i'}}}_{\delta_{ii'}\mu_{kl}+O(\tau^{t+1})}\ket{c_i}\bra{c_{j'}}
\Bigr)\\
&=\rho_Q\sum_{k,l} |\lambda_{kk}|^2(\lambda_{kl}\mu_{kl}^*+\lambda_{kl}^*\mu_{kl})+O(\tau^{t+1})\\
\noalign{\ \hfill\eqref{eq:residual1_corrected}}
\end{alignat*}
which is, up to order
$O(\tau^{t+1})$, proportional to the original state.
\stepcounter{equation}\immediate\write1{\string\newlabel{eq:residual1_corrected}{{\theequation}{\thepage}}}

\end{IEEEproof}

Note that in the proof of Theorem \ref{th:approx}, we have split the
error-operators accordingly based on their expansion
\eqref{eq:serial} in terms of $\sqrt{\tau}$, see
\eqref{eq:def_B_k}--\eqref{eq:def_D_k}.  Clearly, the high order
parts $D_k$ can be completely ignored.  Only the errors $B_k$ of
approximately half the final order have to be corrected
\eqref{eq:correction}, while the errors $C_k$ have to obey some kind
of error detection criterion \eqref{eq:detection}.

\section{Stabilizer and Asymmetric Quantum Codes}
Before presenting our construction of quantum codes tailored to
amplitude damping channels, we investigate the performance of
traditional quantum error-correcting codes on these channels.

Stabilizer codes are a large class of quantum codes which contain
many good quantum codes \cite{thesis:gottesman,nielsenchuang}.  A
stabilizer code $\cQ$ with $n$ qudits encoding $k$ qudits has
distance $d$ if all errors of weight at most $d-1$ can be detected
or have no effect on $\cQ$, and we denote the parameters of $\cQ$ by
$[\![n,k,d]\!]_q$.  Obviously a stabilizer code of distance $2t+1$
corrects $t$ AD error as it corrects $t$ arbitrary errors.

Calderbank-Shor-Steane (CSS) codes \cite{calderbank96,steane96} are a
subclass of the stabilizer codes.  It has been shown that CSS codes
can be used to construct codes for the binary AD channel \cite[Section
  8.7]{thesis:gottesman}. The construction is based on so-called
asymmetric quantum codes, which have a direct generalization to the
qudit case \cite{FJLP03}.  The following theorem shows that those
asymmetric CSS codes can also be used to obtain error correcting codes for qudit
AD channels.
\begin{theorem}\label{thm:CSS}
An $[\![n,k]\!]_q$ CSS code $Q$ with pure $X$-distance $2t+1$ and pure
$Z$-distance $t$ corrects $t$ AD error, i.e., errors up to order
$O(\tau^t)$.
\end{theorem}
\begin{IEEEproof}
The generalized Pauli operators $X_q^kZ_q^l$ form a basis for all
operators on a single qudit. Hence we can expand the error operators
$A_i$ in terms of tensor products of the generalized Pauli operators.
The diagonal error operator $A_0$ of AD channels can be expanded in
terms of the error operators $Z_q^l$, with the expansion coefficients
of the operators $Z_q^l$, $l>0$ being of first order in $\tau$.  The
diagonal of the other error operators $A_{ij}$ or $A_i$ is zero. They
can be expanded in terms of operators $Z_q^l X_q^k$, $k\ne 0$, with
the expansion coefficients being of order $\sqrt{\tau}$.

Note that for CSS codes, $X$ and $Z$ errors can be corrected
independently.  The error operators $B_k$ defined in
\eqref{eq:def_B_k} of the proof of Theorem~\ref{thm:approx} are of
order at most $t/2$ in $\tau$, and hence they contain no more than $t$
$X$-errors and no more than $t/2$ $Z$-errors.  As the code $Q$ has
$X$-distance $2t+1$ and $Z$-distance $t+1$, the error operators $B_k$
can be corrected. Similarly, for the error operators $C_k$ defined in
\eqref{eq:def_C_k}, there are no more than $2t$ $X$-errors and no more
than $t$ $Z$-errors, which can be detected using $Q$.  Hence, using
Theorem~\ref{thm:approx}, it follows that $Q$ corrects all errors up
to order $O(\tau^t)$.
\end{IEEEproof}


\section{Classical Asymmetric Codes}
In this section, we construct quantum codes correcting a single AD
error using classical asymmetric codes.  Codes for the qubit case have
been presented in \cite{LS07,SSS+11}.  Those codes are
self-complementary, i.e., the basis states are of the form
$\ket{\psi_{\mathbf{u}}}=\frac{1}{\sqrt{2}}(\ket{\mathbf{u}}+\ket{\mathbf{\bar{u}}})$,
where $\mathbf{u}$ is an $n$-bit string,
$\mathbf{\bar{u}}=\mathbf{1}\oplus\mathbf{u}$, and $\mathbf{1}$ is the
all-one string.

For the non-binary case with $q>2$, we consider a similar
construction. Define $\bar{X}=X_q^{\otimes n}$, then the basis states are
chosen as
\begin{equation}
\label{eq:sc}
\ket{\psi_{\mathbf{u}}}=\frac{1}{\sqrt{q}}\sum_{l=0}^{q-1}\bar{X}^l\ket{\mathbf{u}}.
\end{equation}
For instance, for $q=3$ and $n=3$, we get
$\ket{\psi_{\mathbf{0}}}=\frac{1}{\sqrt{3}}(\ket{000}+\ket{111}+\ket{222})$.

The quantum code $\mathcal{Q}$ is then spanned by
$\{\ket{\psi_{\mathbf{u}}}\}$, where $\mathbf{u}\in\tilde{C}$ is
some length-$n$ string over the alphabet
$\{0,1,\ldots,q-1\}$ ($\tilde{C}$ is a classical code of length $n$).
The advantage of this construction is that the code automatically
satisfies the error-detection condition for a single $Z_q^l$ error
($l=1,2,\ldots,q-1$), as the code is stabilized by $\bar{X}$. Now
consider a classical code with codewords $C=\{
\mathbf{u}+\alpha\mathbf{1}\colon\mathbf{u}\in\tilde{C},\alpha=0,\ldots,
q-1\}$ and the corresponding quantum code spanned by
$\{\ket{\psi_{\mathbf{u}}}\colon\mathbf{u}\in\tilde{C}\}$.  The
problem of correcting a single error for the qudit AD channels can
then be reduced to finding certain classical codes.

The relevant classical channel is the classical asymmetric channel
\cite{DetCode}.  Let the alphabet be $\mathbb{Z}_q$ with the ordering
$0<1<2< \dots < q-1$. A channel is called asymmetric if any
transmitted symbol $a$ is received as $b\leq a$.  The mostly studied
asymmetric channel, dating back to Varshamov \cite{VAR65}, can be
described by the following asymmetric distance
$\Delta(\mathbf{x},\mathbf{y})$.

\begin{definition}[see \cite{Klo81}]
\label{def:asy}
Let $B=\{0,1,\ldots,q-1\}\subset\mathbb{Z}$.  For
$\mathbf{x},\mathbf{y}\in B^n$, we define
\begin{enumerate}
\item $w(\mathbf{x}):=\sum_{i=1}^n x_i$.
\item $N(\mathbf{x},\mathbf{y}):=\sum_{i=1}^{n}\max\{y_i-x_i,0\}$.
\item
  $\Delta(\mathbf{x},\mathbf{y}):=\max\{N(\mathbf{x},\mathbf{y}),N(\mathbf{y},\mathbf{x})\}$.
\end{enumerate}
\end{definition}
If $\mathbf{x}$ is sent and $\mathbf{y}$ is received, we say that
$w(\mathbf{x}-\mathbf{y})$ errors have occurred (note that $x_i\ge
y_i$ and hence each summand in $w(\mathbf{x}-\mathbf{y})$ is
nonnegative).  A code correcting $t$-errors is called a $t$-code.
\begin{theorem}[see \cite{Klo81}]
A code $C\subset B^n$ corrects $r$ errors of the asymmetrical channel if
and only if $\Delta(\mathbf{x},\mathbf{y})>r$ for all
$\mathbf{x},\mathbf{y}\in C$, $\mathbf{x}\neq\mathbf{y}$.
\end{theorem}

Our goal is to link these classical asymmetric codes to quantum AD
codes. As discussed above, we start from the following definition.
\begin{definition}
A classical code $C$ over the alphabet $B$ is called
self-complementary if for any $\mathbf{x}\in C$,
$\mathbf{1}\oplus\mathbf{x}\in C$.
\end{definition}

For any self-complementary code $C$, there exists another code
$\tilde{C}$ such that $C=\{
\mathbf{u}+\alpha\mathbf{1}\colon\mathbf{u}\in\tilde{C},\alpha=0,\ldots
q-1\}$ and $|C|=q|\tilde{C}|$.  We may, for example, choose all
$\mathbf{u}\in\tilde{C}$ such that the first digit is $0$.  From
$\tilde{C}$ we derive the quantum code $\mathcal{Q}$ spanned by
$\{\ket{\psi_{\mathbf{u}}}\colon\mathbf{u}\in\tilde{C}\}$ as given in
Eq. \eqref{eq:sc}.  Our main result is given by the following theorem.
\begin{theorem}\label{th:main}
If $C$ is a classical (linear or non-linear) self-complementary code
correcting a single error with respect to Definition~\ref{def:asy},
then $\mathcal{Q}$ spanned by
$\{\ket{\psi_{\mathbf{u}}}\colon\mathbf{u}\in\tilde{C}\}$ is a
single-error-correcting code for the qudit AD channels $\Xi$ and
$\mathcal{A}$.
\end{theorem}
\begin{IEEEproof}
Let $E_{ij}=\ket{i}\bra{j}$ with $i,j\in\{0,1,\ldots,q-1\}$ and $i<j$.
For a small decay time $\tau$, in order to improve the fidelity of the
transmission through the qudit AD channel $\mathcal{A}$ given by
Eq. \eqref{eq:Ak} from $1-O(\tau)$ to $1-O(\tau^2)$, it is sufficient to
correct a single $E_{i,i+1}$-error and detect one $Z^l_q$-error for
$l=1,2,\ldots,q-1$. The self-complementary form of
$\ket{\psi_{\mathbf{u}}}$ given in Eq. \eqref{eq:sc} implies that
$\bar{X}\ket{\psi_{\mathbf{u}}}=\ket{\psi_{\mathbf{u}}}$. In turn,
this implies that
$\bra{\psi_{\mathbf{v}}}Z_q^l\ket{\psi_{\mathbf{u}}}=0$ for any
$\mathbf{u},\mathbf{v}$ and $l=1,2,\ldots,q-1$, i.e., the
error-detection condition for a single $Z_q^l$ error is fulfilled.

Next consider a single operator $E_{i,i+1}$.  Every state of the
quantum code is a linear combination of states $\ket{\mathbf{c}}$ with
$\mathbf{c}\in C$.  Applying the operator $E_{i,i+1}$ to
$\ket{\mathbf{c}}$ corresponds to a single asymmetric error.  As the
classical code $C$ corrects a single asymmetric error, the distance
$\Delta(\mathbf{u},\mathbf{v})$ between any two codewords $\mathbf{u}$
and $\mathbf{v}$ is at least two. Therefore, the supports
(set of basis states with non-zero coefficient in the superposition)
of the states $\ket{\psi_{\mathbf{u}}}$ and
$E_{i,i+1}^{(\alpha)}\ket{\psi_{\mathbf{v}}}$ are disjoint for all
positions $\alpha$, where $E_{i,i+1}^{(\alpha)}$ denotes the operator
$E_{i,i+1}$ acting at position $\alpha$.  Hence those states are
mutually orthogonal.  Finally note that for errors $E_{i,i+1}$ acting
on the same position, the operator $E_{i,i+1}^\dagger E_{i,i+1}$ is
diagonal and hence in the span of the operators $Z_q^l$, which can be
detected.
\end{IEEEproof}
\begin{corollary}\label{co:self_complementary}
If there exists an $(n,K,3)_q$ self-com\-ple\-men\-tary code $C$,
then there exists an $(\!(n,K/q)\!)_q$ quantum code correcting a
single AD error.
\end{corollary}

Such codes have, e.\,g., been studied in \cite{EG13}. For linear codes, we have the following corollary.

\begin{corollary}
\label{co:CSS}
If there exists an $[n,k+1,3]_q$ linear code $C$ containing
the all-one-vector $\mathbf{1}\in C$, then there exists an
$[\![n,k]\!]_q$ CSS code correcting a single AD error.
\end{corollary}
In the preceding corollaries we have used the notation $(n,K,d)_q$ for
a classical code of length $n$ with $K$ codewords and minimum distance
$d$ over an alphabet with $q$ elements, and the notation
$[n,k,d]_q=(n,K=q^k,d)$ when the code is linear.

\section{Single-Error-Correcting Codes: Even Lengths}
\label{sec:even}

We now use Theorem \ref{th:main} to construct some families of good
single-error-correcting AD codes. For this, we need to find some good
self-complementary single-error-correcting classical asymmetric codes.
The best known direct construction of single-error-correcting
codes for the binary asymmetric channel is the so-called
Varshamov-Tenengolts (VT)-Constantin-Rao (CR) code ~\cite{VT65,CR79},
with a natural generalization to $q>2$. These VT-CR codes are
non-linear codes, in both the binary and non-binary cases.

For the binary case, many of these VT-CR codes are indeed
self-complementary, and so they can be used to construct families of
good single-error-correcting quantum AD codes \cite{SSS+11}. As the
VT-CR codes are nonlinear, the corresponding quantum codes are
nonadditive codes.  Unfortunately, for the non-binary case, the VT-CR
codes are no longer self-complementary, so one needs some other
constructions of good single-error-correcting for asymmetric channels.

We will use the idea of generalized concatenation, which has been
discussed in the context of constructing binary AD codes in
\cite{SSS+11}, and in the context of constructing (classical)
asymmetric codes in \cite{GSS+15}. This method will allow us to
construct good self-complementary asymmetric linear codes for the
non-binary case, which will lead to good single-error-correcting
quantum codes for AD channels.

\subsection{Qutrit Codes}

First, we consider the case of $q=3$.  For the generalized
concatenation construction, we choose the outer code as some ternary
classical code over the alphabet $\{\tilde{0}, \tilde{1},
\tilde{2}\}$, and the inner codes as:
\begin{equation}
\label{eq:ternary}
C_{\tilde{0}}=\{00,11,22\},\ C_{\tilde{1}}=\{01,12,20\},\ C_{\tilde{2}}=\{02,10,21\}.
\end{equation}

Then we have the following result.
\begin{theorem}\label{th:ADXi}
For $n$ even, generalized concatenation with an outer $[n/2,k,3]_3$
code results in an $[n,n/2+k]_3$ self-complementary linear code $C$.
This code leads to an $[\![n,n/2+k-1]\!]_3$ quantum stabilizer code
$\mathcal{Q}$, correcting a single error for the channels $\Xi$ and
$\mathcal{A}$.
\end{theorem}
\begin{IEEEproof}
Note that $C_{\tilde{0}}$, $C_{\tilde{1}}$, and $C_{\tilde{2}}$ are
all self-com\-ple\-men\-tary codes correcting a single asymmetric
error.  Therefore, any outer ternary code will lead to a
self-complementary ternary code $C$, and hence a quantum code
$\mathcal{Q}$.  A single amplitude damping error induces only a single
error with respect to $\tilde{0}$, $\tilde{1}$, $\tilde{2}$.  As the
outer ternary code has distance $3$, such an error can be corrected.
\end{IEEEproof}

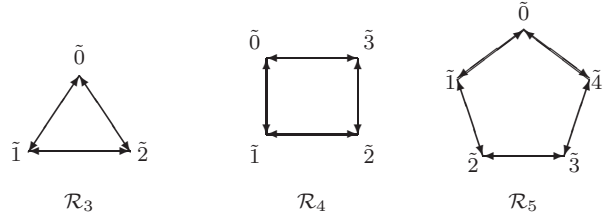
\begin{figure}[h!]
\vskip-\baselineskip
\centerline{\footnotesize\unitlength1.6\unitlength%
\begin{tabular}{ccc}
\begin{picture}(50,50)(0,5)
\put(25,33){\makebox(0,0){$\tilde{0}$}}
\put(10,10){\makebox(0,0){$\tilde{1}$}}
\put(40,10){\makebox(0,0){$\tilde{2}$}} \put(37,10){\vector(-1,0){24}}
\put(13,10){\vector(1,0){24}} \put(13,10){\vector(2,3){12}}
\put(25,28){\vector(-2,-3){12}} \put(37,10){\vector(-2,3){12}}
\put(25,28){\vector(2,-3){12}}
\end{picture}
&\unitlength0.9\unitlength
\begin{picture}(50,50)(0,5)
\put(10,40){\makebox(0,0){$\tilde{0}$}}
\put(10,10){\makebox(0,0){$\tilde{1}$}}
\put(40,10){\makebox(0,0){$\tilde{2}$}}
\put(40,40){\makebox(0,0){$\tilde{3}$}} \put(13,15){\vector(1,0){24}}
\put(37,15){\vector(-1,0){24}} \put(13,35){\vector(1,0){24}}
\put(37,35){\vector(-1,0){24}} \put(13,15){\vector(0,1){20}}
\put(13,35){\vector(0,-1){20}} \put(37,15){\vector(0,1){20}}
\put(37,35){\vector(0,-1){20}}
\end{picture}
&\unitlength0.8\unitlength
\begin{picture}(50,50)(0,5)
\put(25,50){\makebox(0,0)[b]{$\tilde{0}$}}
\put(5,32.5){\makebox(0,0)[r]{$\tilde{1}$}}
\put(10,8){\makebox(0,0){$\tilde{2}$}}
\put(40,8){\makebox(0,0){$\tilde{3}$}}
\put(45,32.5){\makebox(0,0)[l]{$\tilde{4}$}}
\put(37,10){\vector(-1,0){24}} \put(13,10){\vector(1,0){24}}
\put(13,10){\vector(-1,3){7.5}} \put(5.5,32.5){\vector(1,-3){7.5}}
\put(37,10){\vector(1,3){7.5}} \put(44.5,32.5){\vector(-1,-3){7.5}}
\put(5.5,32.5){\vector(4,3){19.5}} \put(25,47.5){\vector(-4,-3){19.5}}
\put(25,47.5){\vector(4,-3){19.5}}
\put(44.5,32.5){\vector(-4,3){19.5}}
\end{picture}
\\ $\mathcal{R}_3$ & $\mathcal{R}_4$ & $\mathcal{R}_5$
\end{tabular}}
\caption{The induced channel $\mathcal{R}_3$ for $q=3$ (which is just
  the ternary symmetric channel), the induced channel $\mathcal{R}_4$
  for $q=4$, and the induced channel $\mathcal{R}_5$ for $q=5$. The
  arrows indicate the possible transitions between symbols.}
\label{fig:channel2}
\end{figure}

Note that with respect to the symbols $\tilde{0},\tilde{1},\tilde{2}$,
the induced channel $\mathcal{R}_3$ is nothing but the ternary
symmetric channel shown in Fig.~\ref{fig:channel2}.

\begin{example}
For $n=6$, take the outer code of length $n/2=3$ as
$\{\tilde{0}\tilde{0}\tilde{0}, \tilde{1}\tilde{1}\tilde{1},
\tilde{2}\tilde{2}\tilde{2}\}$ with distance $3$.  Generalized
concatenation yields a self-complementary ternary linear code of
dimension $4$.  The corresponding quantum code $\mathcal{Q}$ encodes
$6/2+1-1=3$ qutrits.  Both the best corresponding
single-error-correcting quantum code $[\![6,2,3]\!]_3$ and the best
possible asymmetric CSS code $[\![6,2,\{3,2\}]\!]_3$ (see Corollary
\ref{co:CSS}) encode only $2$ qutrits.
\end{example}



\subsection{The Case $q>3$ }

For $q=4$, we choose the inner codes as
\begin{alignat}{5}
\label{eq:quad2}
C_{\tilde{0}}&=\{00,11,22,33\},\quad&C_{\tilde{1}}&=\{01,12,23,30\},
\nonumber\\ C_{\tilde{2}}&=\{02,13,20,31\},&C_{\tilde{3}}&=\{03,10,21,32\}.
\end{alignat}
Similar as in Theorem \ref{th:ADXi}, an outer code with distance three
yields a self-complementary code from which a quantum AD code can be
derived.  However, in this case, the induced channel for the outer
code is no longer symmetric.  A single damping error will, for
example, never map a codeword of the inner code $C_{\tilde{0}}$ to a
codeword of $C_{\tilde{2}}$.  So on the level of the outer code, there
are no transitions between $\tilde{0}$ and $\tilde{2}$, or between
$\tilde{1}$ and $\tilde{3}$.  The induced quaternary channel
$\mathcal{R}_4$ is shown in Fig.~\ref{fig:channel2}, where we see that
errors only happen between `neighbors.'

The above constructions for $q=3,4$ have a direct generalization to
general $q>2$.  For a given $q$, choose the outer code as some code
over the alphabet $\{\tilde{0}, \tilde{1}, \ldots,\widetilde{q-1}\}$.
The $q$ inner codes
$C_{\tilde{0}},C_{\tilde{1}},\ldots,C_{\widetilde{q-1}}$ are the
double-repetition code $C_{\tilde{0}}=\{00,11,\ldots,(q-1)(q-1)\}$ and
all its $q-1$ cosets $C_{\tilde{i}}=C_{\tilde{0}}\oplus (0i)$, i.e.,
we apply the rule that $0i\in C_{\tilde{i}}$.  It is straightforward
to check that each inner code has asymmetric distance $2$, hence
corrects a single asymmetric error.  Similar as in the case of $q=4$,
a single damping error will only drive transitions between
$\tilde{i},\tilde{j}$ for $\tilde{i}=\tilde{j}\pm \tilde{1}$. For
instance, for $q=5$, the induced channel $\mathcal{R}_5$ is shown in
Fig.~\ref{fig:channel2}.  In general, we will write the induced
channel as $\mathcal{R}_q$ for outer codes over $\{\tilde{0},
\tilde{1}, \ldots,\widetilde{q-1}\}$.

Similar as Theorem~\ref{th:ADXi}, in general we have the following
theorem.
\begin{theorem}
\label{th:nonbi1}
For $n$ even, an outer $[n/2,k]_q$ code correcting a single error for
the channel $\mathcal{R}_q$ leads to an $[n,n/2+k]_q$
self-complementary linear code $C$ and hence an $[\![n,n/2+k-1]\!]_q$
quantum code $\mathcal{Q}$, correcting a single error for the qudit AD
channels $\Xi$ and $\mathcal{A}$.
\end{theorem}

Note that the channel $\mathcal{R}_q$ is no longer a symmetric
channel, so outer codes of Hamming distance $3$ are no longer expected
to give the best codes.  It turns out, however, that
single-error-correcting codes for the channel $\mathcal{R}_q$ are
equivalent to single-symmetric-error correcting codes in Lee
metric~\cite{Ber68} (see also~\cite{KBE11}), for which optimal linear
codes are known (for a more detailed discussion, see~\cite{GSS+15}).

\section{Single-Error-Correcting Codes: Odd Lengths}

The construction of AD codes for even lengths given in
Sec.~\ref{sec:even} based on generalized concatenation is relatively
straightforward. The inner codes are just $1$-codes of length $2$ with
$q$ codewords and their cosets.  In \cite{GSS+15}, codes of odd length
were obtained using a mixed-alphabet code, treating one position
differently.  This does not directly translate to the situation
considered here, as the resulting code has to be self-complementary.

Instead, we will use different inner codes, one of odd lengths and the
length-two code from above. In particular, we can directly search for $q$
mutually disjoint inner codes of length $3$ which are $1$-codes.

For $q=4$, consider the following $\mathbb{Z}_4$-linear code $C_{0'}$
of length $3$ generated by $\{111,002,020\}$:
\begin{alignat}{7}
000\ 111\ 222\ 333&\qquad
002\ 113\ 220\ 331\nonumber\\ 020\ 131\ 202\ 313&\qquad
022\ 133\ 200\ 311.
\end{alignat}
The code $C_{0'}$ has asymmetric distance $2$, as well as the three
cosets $C_{1'}=C_{0'}+001$, $C_{2'}=C_{0'}+010$, and
$C_{3'}=C_{0'}+100$.  Applying generalized concatenation to the outer
code
$\{\tilde{0}\tilde{0}0',\tilde{1}\tilde{1}1',\tilde{2}\tilde{2}2',\tilde{3}\tilde{3}3'\}$
and the inner codes of length $2$ and $3$ for the first two and the
third position, respectively, yields a self-complementary $1$-code
$[7,5]_4$.  The corresponding quantum code has parameters
$[\![7,4]\!]_4$.

Note that the induced channel on the alphabet $\{0',1',2',3'\}$ is no
longer $\mathcal{R}_4$, but the symmetric channel over
$\mathbb{Z}_4$. Therefore we have the following theorem for $q=4$.
\begin{theorem}
\label{th:oddq4}
For $n$ odd, an outer $[(n-1)/2,k,3]_4$ code leads to an
$[n,(n+1)/2+k]_4$ self-complementary linear $1$-code $C$.  The
resulting quantum code $\mathcal{Q}=[\![n,(n-1)/2+k]\!]_4$ corrects a
single error for the qudit AD channels $\Xi$ and $\mathcal{A}$.
\end{theorem}
\begin{IEEEproof}
The inner codes $C_{\tilde{i}}$ of length two as well as the inner
codes $C_{i'}$ of length three are self-complementary $1$-codes.  The
outer code has distance $3$ which ensures that a single error mixing
the inner codes can be corrected.  For the outer code, we always take
the last coordinate to be of type $s'$, and all the other coordinates
to be of type $\tilde{s}$, for $s=0,\ldots,3$.  Hence, the inner code
for the last coordinate of the outer code has length three, while the
other inner codes have length two. Therefore, for an outer
$[(n-1)/2,k,3]_4$ linear code, generalized concatenation yields an
$[n,(n+1)/2+k]_4$ self-complementary linear $1$-code $C$,
corresponding to an $[\![n,(n-1)/2+k]\!]_4$ quantum code.
\end{IEEEproof}

We emphasize that the construction related to Theorem \ref{th:oddq4}
is valid only for $q=4$.  For $q>5$, the $\mathbb{Z}_q$-linear code
$C_0$ generated by $\{111,013\}$ and its $q$ cosets are all
self-complementary codes with asymmetric distance $2$.  For this, note
that $\Delta(\mathbf{x},\mathbf{y})=1$ if and only if, up to
permutation, $\mathbf{x}-\mathbf{y}\in\{(1,0,0),(1,-1,0)\}$. For
$q>5$, the code $C_0$ does not contain such a vector.  Hence we obtain
the analogue result as in Theorem \ref{th:oddq4} for $q>5$.

For $q=3$ and $q=5$, however, we cannot partition the trivial code
$[3,3]_q$ into $q$ self-complementary codes $[3,2]_q$ with asymmetric
distance $2$, substantiated by exhaustive search. However, we can use
an inner code of length five, resulting in the following theorem.
\begin{theorem}
\label{th:oddl}
For $q=3,5$ and $n$ odd, an outer $[(n-3)/2,k]_q$ code correcting a
single error for the symmetric channel leads to an $[n,(n+1)/2+k]_q$
self-complementary code $\mathcal{C}$ and hence an $[\![n,(n-1)/2+k]\!]_q$
quantum code $\mathcal{Q}$, correcting a single error for the qudit AD
channel $\mathcal{A}$ for $q=3,5$.
\end{theorem}

\begin{IEEEproof}
We can map the first digit of the outer code to $q$
groups of codes of length $5$.
Again by exhaustive search, we find that for $q = 3,5$, we cannot
partition $[5,5]_q$ into $q$ self-complementary codes $[5, 4]_q$ with
asymmetric distance $2$, but we can get $q$ codes $[5,3]_q$. For
$q=3$, the code $C_{0'}$ is generated by $\{00011,01201,11111\}$,
while for $q=5$ it is generated by $\{00011,00102,11111\}$.

Then our construction can be described as follows. For an
$[(n-3)/2,k]_q$ outer code, we use the length-5 code described above
as the inner code for the first digit, and use length-2 code for the
remaining $(n-5)/2$ digits, leading to a code with length $1 \times 5
+ \frac{n-5}{2} \times 2 = n$. Similar as in the proof of
Theorem~\ref{th:oddq4}, it follows that the resulting code corrects a
single AD error.
\end{IEEEproof}

In the nonlinear case, we can find larger codes.  The results are
summarized in Theorems \ref{th:oddnl3}, \ref{th:oddnl51}, and
\ref{th:oddnl52}.


\begin{theorem}
\label{th:oddnl3}
For $q=3$ and $n$ odd, an outer $\bigl((n-3)/2,K,3\bigr)_3$ code leads to an
$(n,33\times3^{(n-5)/2} K)_3$ self-complementary code. The resulting
$(\!(n,11\times3^{(n-5)/2} K)\!)_3$ quantum code corrects one error for
the AD channels $\Xi$ and $\cA$.
\end{theorem}
\begin{IEEEproof}
An exhaustive search reveals that for $q=3$, we can find three
disjoint self-complementary codes of length $5$ and asymmetric
distance $2$ with at most $33$ codewords.  Let
\begin{alignat}{5}
\tilde{C}_{0'} = \{00000, 00011, 00112, 00220, 01021,
01110,&\nonumber \\ 01202, 02022, 02101, 02120, 02211& \}
\end{alignat}
Then $C_{0'}=\{\mathbf{u}+\alpha\mathbf{1}\mid\mathbf{u}\in
\tilde{C}_{0'},\alpha=0,1,2\}$, $C_{1'}=\{\mathbf{u} +
00001\mid\mathbf{u} \in C_{0'}\}$, $C_{2'}=\{\mathbf{u} +
00002\mid\mathbf{u} \in C_{0'}\}$.  We construct the code similarly as
in Theorem~\ref{th:oddl}, i.e,. the inner code for the fist digit is
$C_{i'}$, and for the remaining $(n-5)/2$ digits the inner code is
$C_{\tilde i}$.
\end{IEEEproof}

For $q=5$, we consider two constructions based on nonlinear codes.
\begin{theorem}
\label{th:oddnl51}
Then for $q=5$ and $n$ odd, an outer $\bigl((n-1)/2, K, 3\bigr)_5$ code leads to
an $(n,20\times 5^{(n-3)/2}K)_5$ self-complementary code. The
resulting $(\!(n, 4 \times 5^{(n-3)/2}K)\!)_5$ quantum code corrects one
error for the AD channel $\cA$.
\end{theorem}
\begin{IEEEproof}
 Although we cannot partition $[3,3]_5$, a weaker result can be
obtained. Let
\begin{alignat*}{5}
\tilde{C}_{0'}&=\{000, 002, 020, 022\},\quad&
\tilde{C}_{1'}&=\{001, 004, 021, 024\},\\
\tilde{C}_{2'}&=\{003, 011, 031, 033\},&
\tilde{C}_{3'}&=\{010, 023, 041, 043\},\\
\tilde{C}_{ 4'}&=\{012, 014, 032, 034\}.
\end{alignat*}
Furthermore, set $C_{i'}=\{\mathbf{u}+
\alpha\mathbf{1}\mid\mathbf{u}\in \tilde{C}_{i'},
\alpha=0,1,2,3,4\}$.

Then we have a construction similar to Theorem~\ref{th:oddq4}, i.e.,
we use one copy of the length-3 inner code and $(n-3)/2$ copies of the
length-2 inner code.
\end{IEEEproof}

\begin{theorem}
\label{th:oddnl52}
For $q=5$ and $n$ odd, an outer $((n-3)/2, K, 3)_5$ code leads to an
$(n,295\times 5^{(n-5)/2}K)_5$ self-complementary code. The resulting
$(\!(n, 59 \times 5^{(n-5)/2}K)\!)_5$ quantum code can correct one error
for the AD channel $\cA$.
\end{theorem}
\begin{IEEEproof}
In this construction we use self-complementary codes of length $5$. By
non-exhaustive search, we can find a self-complementary code with $295$ codewords.
The $59$ codewords of $\tilde{C}_{0}'$ are shown in
Table~\ref{table:q5}.  From those we derive $C_{0'}=\{\mathbf{u} +
\alpha\mathbf{1}\mid\mathbf{u}\in\tilde{C}_{0'},\alpha=0,1,2,3,4\}$
and $C_{i'}=\{\mathbf{u} + 0000i\mid\mathbf{u}\in C_{0'}\}$, $1
\leq i < 5$.
\begin{table}[ht!]
\centering
\caption{Code Construction for $q=5$}
\begin{tabular}{|ccccc|}
 \hline
  00000 & 00202 & 01241 & 02200 & 03110\\
  00002 & 00220 & 01404 & 02203 & 03212\\
  00013 & 00223 & 01412 & 02211 & 03231\\
  00020 & 00244 & 02000 & 02223 & 03233\\
  00031 & 00303 & 02002 & 02314 & 03300\\
  00033 & 00311 & 02013 & 02321 & 03303\\
  00044 & 00314 & 02021 & 02332 & 03342\\
  00111 & 00330 & 02032 & 02424 & 03410\\
  00114 & 00332 & 02034 & 02440 & 03412\\
  00122 & 00424 & 02114 & 03041 & 03431\\
  00141 & 00442 & 02130 & 03044 & 04234\\
  00200 & 01133 & 02143 & 03102 & \\
 \hline
\end{tabular}
\label{table:q5}
\end{table}
Then we have a construction similar to that in Theorem~\ref{th:oddl}.
\end{IEEEproof}

As shown in Table \ref{table:qutrit}, for many lengths, the
construction based on Theorems \ref{th:ADXi}, \ref{th:oddl}, and
\ref{th:oddnl3} outperforms both the best known quantum codes with
distance $3$, and the CSS codes of Corollary \ref{co:CSS}.  The
dimension of the asymmetric quantum codes (AQECC) is taken from
\cite{EG13}. Only when $n = 13$, our construction performs worse than
CSS and AQECC. This might be due to the fact that the outer code we
use here is the CSS code $[\![5,1,\{3, 2\}]\!]$, which is not efficient
since we also have CSS code $[\![4,1,\{3, 2\}]\!]$.
\begin{table}[h!]
\caption{dimension of single-error-correcting quantum AD
codes from the $GF(3^2)$ construction with distance $3$, the
CSS construction, asymmetric quantum codes (AQECC), and the
generalized concatenation construction (GC).}
\centering
\begin{tabular}{|c|c c c c c|}
 \hline
 $n$ & $GF(3^2)$ & CSS & AQECC & GC (linear) & GC (nonlinear)\\
 \hline
4 & $3^0$ & $3^0$ & 1 & 3 &3   \\
5 & $3^1$ & $3^1$ & 6 & $3^2$ & 11\\
6 & $3^2$ & $3^2$ & 11 & $3^3$ & $3^3$  \\
7 & $3^3$ & $3^3$ & 29 & $3^3$ & $11 \times 3^1$\\
8 & $3^4$ & $3^4$ & 84 & $3^5$ & $3^5$  \\
9 & $3^5$ & $3^5$ & $3^5$ &  $3^5$ & $11 \times 3^3$ \\
10 & $3^6$ & $3^6$ & $3^6$ & $3^6$ & $3^6$  \\
11 & $3^6$ & $3^7$ & $3^7$ &  $3^7$ & $11 \times 3^5$\\
12 & $3^7$ & $3^8$ & $3^8$ & $3^8$ & $3^8$  \\
13 & $3^8$ & $3^9$ & $3^9$ &  $3^8$ & $11 \times 3^6$\\
14 & $3^9$ & $3^9$ & $3^9$ & $3^{10}$ & $3^{10}$ \\
15 & $3^{10}$ & $3^{10}$ &  $3^{10}$ & {$3^{10}$} & $11 \times 3^8$\\
16 & $3^{11}$ & $3^{11}$ & $3^{11}$ & $3^{12}$ & $3^{12}$  \\
 \hline
\end{tabular}
\label{table:qutrit}
\end{table}

\section{Multi-Error-Correcting Codes}

For the binary case, multi-error-correcting amplitude damping codes
are discussed in~\cite{DGJZ10}. The basic idea is that for the
encoding $\ket{0_L}=\ket{01}$, $\ket{1_L}=\ket{10}$, the amplitude
damping channel simulates a binary erasure channel. So one can use
erasure-correcting code as outer codes to build codes correcting
amplitude damping errors.  In this section we consider generalizations
of this construction, for both the binary and non-binary cases.

It is mentioned in~\cite{DGJZ10} that a possible generalization is to
use $\ket{001},\ket{010},\ket{100}$ as the inner code, and a distance
$t+1$ quantum code as an outer code. However, it turns out that one
can actually use $\ket{001},\ket{010},\ket{100},\ket{111}$ as the
inner code, and a distance $t+1$ quantum code as an outer code.  Here
a single damping error will be treated as an erasure, and two damping
errors or no damping can be treated as an error which is taken care of
by the outer code.

To generalize this idea to the case $q>2$, one can take a similar
approach.  In this section we consider the channel $\mathcal{A}$
with Kraus operators given by Eq.~\eqref{eq:Ak}.  For $q=3$, one can
take the encoding $\ket{0_L}=\ket{11}$, $\ket{1_L}=\ket{02}$,
$\ket{2_L}=\ket{20}$. Then the amplitude damping channel simulates a
ternary erasure channel. So one can use erasure-correcting codes as
outer codes to build codes correcting amplitude damping errors.
Actually, using a similar idea as the construction based on
$\ket{001}$, $\ket{010}$, $\ket{100}$, $\ket{111}$ for the binary
case, one can use $\ket{0_L}=\ket{00}$, $\ket{1_L}=\ket{20}$,
$\ket{2_L}=\ket{11}$, $\ket{3_L}=\ket{02}$, and $\ket{4_L} =
\ket{22}$ as the inner code, and still a distance $t+1$ quantum code
as an outer code.  Here we give the general construction.

Assume the length of the inner code is $m$. We choose the set
\begin{equation}
S=\{\ket{a_1a_2\ldots a_m}\mid\text{$a_1+a_2+\ldots+a_m$ is even}\}
\end{equation}
as the orthonormal basis of the code, and let
\begin{equation}
K = |S|. \label{defK}
\end{equation}
If we have an outer code $[\![n,k,t+1]\!]_K$, we get an $[\![nm,K^k]\!]_q$
code correcting $t$ errors for $\cA$.  For $q=3$, the code also
corrects errors for the qutrit channel $\Xi$.
We have the following theorem:
\begin{theorem}
\label{th:mec} The code constructed above corrects $t$ errors for
the channels $\cA$ and $\Xi$.
\end{theorem}
\begin{IEEEproof}
We want to prove that the condition in Theorem \ref{th:approx}
holds. Let $\cQ$ be the outer $[\![n, k, t + 1]\!]_K$ code. Take any
two error operators $E_{\ell}$ and $E_{\ell'}$, and for any two vectors
$\ket{\psi_i}$ and $\ket{\psi_j}$, we consider $\bra{\psi_i}E_{\ell}^\dag
E_{\ell'}\ket{\psi_j}$. Suppose that on some inner code, $E_{\ell}$ has an odd
number of errors while $E_{\ell'}$ has an even number of errors.  Then
$E_{\ell'}\ket{\psi_j}$ will be in the space spanned by $S$ while
$E_{\ell}\ket{\psi_i}$ will be in the space perpendicular to $S$. So
$\bra{\psi_i}E_{\ell}^\dag E_{\ell'}\ket{\psi_j}=0$ and the condition in
Theorem \ref{th:approx} automatically holds. By symmetry, this
argument also holds when $E_{\ell}$ is even and $E_{\ell'}$ is odd.  Now assume
that on each inner code, the number of errors corresponding to $E_{\ell}$
and $E_{\ell'}$ have the same parity.  We consider the series
expansion of the operator $E_{\ell}^\dag E_{\ell'}$ with respect to
$\tau$.  For this, we expand the tensor factor acting on the $i$-th
particle such that $A_{ij}$ corresponds to the term of order
$\tau^j$. Combined, we get
\begin{align}
E_{\ell}^\dag E_{\ell'} &= (A_{10} + A_{11}\tau + A_{12}\tau^2 + \ldots) \nonumber \\
& \otimes (A_{20} + A_{21}\tau + A_{22}\tau^2 + \ldots) \nonumber \\
& \otimes \ldots \otimes (A_{m0} + A_{m1}\tau + A_{m2}\tau^2 + \ldots)
\end{align}
Note that each of $A_{10}, A_{20}, \ldots A_{m0}$ is either the
identity operator or the zero operator, which can be proved if we take
the limit $\tau \to 0$. For the terms of the form $\Omega(\tau^t)$,
there are at most $t$ non-identity terms on the inner codes, which
will become $0$ according to the error-detection criterion. Thus the
condition in Theorem \ref{th:approx} is satisfied.

\end{IEEEproof}

Finally we give an explicit expression for the dimension of the code
$K$ defined in \eqref{defK} in terms of $q$ and $m$.
\begin{theorem}
\begin{equation}
K = \left\{
\begin{tabular}{ll}
  $q^m / 2$ & if $q$ is even. \\
  $(q^m + 1) / 2$ & if $q$ is odd.
\end{tabular}
\right.
\end{equation}
\end{theorem}
\begin{IEEEproof}
When $q$ is even, we map $\ket{a_1a_2\ldots a_m}$ to
$\ket{q-1-a_1,a_2\ldots a_m}$, which is a one-to-one mapping from $S$
to $\bar S$.  When $q$ is odd, let $i=\min\{j\colon a_j\not =0\}$, and we
map $\ket{a_1a_2\ldots a_m}$ to $\ket{a_1\ldots
  a_{i-1},q-a_i,a_{i+1}\ldots a_m}$, which is a one-to-one mapping
from $S\setminus\{\ket{00\ldots 0}\}$ to $\bar S$. So $|S|=|\bar S|$
when $q$ is even, and $|S|=|\bar S|+1$ when $q$ is odd. Thus
$|S|=q^m/2$ when $q$ is even, and $|S|=(q^m+1)/2$ when $q$ is odd.
\end{IEEEproof}
In Table~\ref{table:MEC}, we compare our codes with both stabilizer
codes and AQECCs.  We use the construction with $q = 3$ and $m = 2$,
and compare the codes that corrects $t$ errors. From
$[\![n/2,k,t+1]\!]_5$ codes given (see also \cite{LX09}), we can
construct our $(\!(n,K)\!)_3$ codes where $K = 5^k$, and we compare
them with stabilizer codes $[\![n,k',2t+1]]_3$ that have dimension
$K'=3^{k'}$.  It can be seen that our construction outperforms
stabilizer codes, and our performance gets better in comparison for
larger $t$.  In most cases, our codes are better also better than the
asymmetric CSS codes.  Some of the CSS codes are optimal codes
taken from \cite{FJLP03}, and others are based on the best known
classical ternary codes \cite{Grassl:codetables}.  For comparison, we
also list an upper bound $k''_{\text{max}}$ on the maximal dimension
of an AQECC based on the known bounds for classical codes. In most
cases, this bound cannot be achieved.
\begin{table}[h!]
\caption{Dimension of quantum codes from our construction that correct
  $t$ AD errors compared with stabilizer codes $[\![n,k',2t+1]\!]_3$
  and pure asymmetric CSS codes with parameters $[\![n,k'',\{2t+1,t+1\}]\!]_3$.}
\centering
\begin{tabular}{c c c c c l c c}
 \hline
 $t$ & $n$ & $K$ & $\log_3 K$ & $k'$ & $k''$ &$k''_{\text{max}}$\\
 \hline
 2 & 10 & 5     & 1.465 & 1 & $[\![10,1,\{5,3\}]\!]_3$ & 2\\
 2 & 12 & 25    & 2.930 & 2 & $[\![12,3,\{5,3\}]\!]_3$ & 3\\
 2 & 14 & 125   & 4.395 & 4 & $[\![14,4,\{5,3\}]\!]_3$ & 4\\
 2 & 16 & 625   & 5.860 & 5 & $[\![16,5,\{5,3\}]\!]_3$ & 5\\
 2 & 18 & 3125  & 7.325 & 6 & $[\![18,7,\{5,3\}]\!]_3$ & 7\\
 2 & 20 & 15625 & 8.790 & 8 & $[\![20,9,\{5,3\}]\!]_3$ & 9\\
 \hline
 3 & 14 & 5     & 1.465 & N/A & $[\![14,0,\{7,4\}]\!]_3$ & 0\\
 3 & 16 & 25    & 2.930 & 0   & $[\![16,1,\{7,4\}]\!]_3$ & 1\\
 3 & 18 & 125   & 4.395 & 1   & $[\![18,3,\{7,4\}]\!]_3$ & 3\\
 3 & 20 & 625   & 5.860 & 3   & $[\![20,4,\{7,4\}]\!]_3$ & 5\\
 3 & 22 & 3125  & 7.325 & 4   & $[\![22,6,\{7,4\}]\!]_3$ & 6\\
 3 & 24 & 15625 & 8.790 & 6   & $[\![24,6,\{7,4\}]\!]_3$ & 8\\
 \hline
 4 & 18 & 5     & 1.465 & N/A & N/A & --\\
 4 & 20 & 25    & 2.930 & N/A & $[\![20,0,\{9,5\}]\!]_3$ & 1\\
 4 & 24 & 625   & 5.860 & 1 & $[\![24,4,\{9,5\}]\!]_3$ & 5\\
 4 & 26 & 3125  & 7.325 & 2 & $[\![26,4,\{9,5\}]\!]_3$ & 6\\
 4 & 28 & 15625 & 8.790 & 3 & $[\![28,5,\{9,5\}]\!]_3$ & 8\\
 \hline
 5 & 26 & 5     & 1.465 & N/A & $[\![26,0,\{13,6\}]\!]_3$ &3\\
 5 & 28 & 25    & 2.930 & 0 & $[\![28,1,\{11,6\}]\!]_3$ & 5 \\
 5 & 30 & 125   & 4.395 & 1 & $[\![30,2,\{11,6\}]\!]_3$ & 6\\
 5 & 32 & 625   & 5.860 & 1 & $[\![32,4,\{11,6\}]\!]_3$ & 7\\
 5 & 38 & 3125  & 7.325 & 4 & $[\![38,7,\{11,6\}]\!]_3$ & 12\\
 5 & 40 & 15625 & 8.790 & 6 & $[\![40,8,\{12,6\}]\!]_3$ & 14\\
 \hline
 6 & 30 & 5     & 1.465 & N/A & ? & 0 \\
 6 & 36 & 25    & 2.930 & 0 & ? & 6\\
 6 & 38 & 125   & 4.395 & 1 & $[\![38,2,\{13,7\}]\!]_3$ & 8\\
 6 & 40 & 625   & 5.860 & 1 & $[\![40,3,\{13,7\}]\!]_3$ & 10\\
 6 & 42 & 3125  & 7.325 & 2 & $[\![42,5,\{13,7\}]\!]_3$ & 11\\
 6 & 44 & 15625 & 8.790 & 2 & $[\![44,6,\{13,7\}]\!]_3$ & 13\\
 \hline
\end{tabular}
\label{table:MEC}
\end{table}

\section{Applications to other AD channels}
Our method can also be applied to some other amplitude damping
processes. For instance, when $q=3$, in addition to the channel $\Xi$,
there are the following natural decay processes of three-level atoms:
the $\Lambda$-pattern or the $V$ pattern. They are illustrated
together with the $\Xi$ channel in Fig.\ref{fig:3levelatom}:
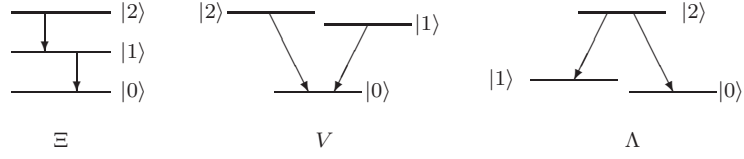
\begin{figure}[h!]
\vskip-\baselineskip
\centerline{\footnotesize\unitlength0.75\unitlength%
\begin{tabular}{c@{\kern10mm}c@{\kern10mm}c}
\begin{picture}(70,90)(0,0)
\put(10,10){\line(1,0){50}}
\put(10,30){\line(1,0){50}}
\put(10,50){\line(1,0){50}}
\put(27,50){\vector(0,-1){20}}
\put(43,30){\vector(0,-1){20}}
\put(65,10){\makebox(0,0)[l]{$\ket{0}$}}
\put(65,30){\makebox(0,0)[l]{$\ket{1}$}}
\put(65,50){\makebox(0,0)[l]{$\ket{2}$}}
\end{picture}
&
\begin{picture}(120,70)(0,0)
\put(11,50){\line(1,0){44}}
\put(60,44){\line(1,0){44}}
\put(35,10){\line(1,0){44}}
\put(32,50){\vector(1,-2){20}}
\put(82,44){\vector(-1,-2){17}}
\put(3,50){\makebox(0,0){$\ket{2}$}}
\put(112,44){\makebox(0,0){$\ket{1}$}}
\put(87,10){\makebox(0,0){$\ket{0}$}}
\end{picture}
&
\begin{picture}(115,75)(10,10)
\put(066,20){\line(1,0){44}}
\put(016,26){\line(1,0){44}}
\put(040,60){\line(1,0){44}}
\put(068,60){\vector(1,-2){20}}
\put(055,60){\vector(-1,-2){17}}
\put(117,20){\makebox(0,0)[1]{$\ket{0}$}}
\put(008,26){\makebox(0,0)[r]{$\ket{1}$}}
\put(092,60){\makebox(0,0)[l]{$\ket{2}$}}
\end{picture}
\\
$\Xi$
&
$V$
&
$\Lambda$
\end{tabular}}
\caption{Decay processes for three-level atoms with different level-structures.}
\label{fig:3levelatom}
\end{figure}

For the $\Lambda$-pattern, the master
equation is
\begin{alignat*}{5}\label{eq:master2}
\frac{d\rho}{d\tau}={}&k_1(2\sigma_{12}^{-}\rho\sigma_{12}^{+}-\sigma_{12}^{+}\sigma_{12}^{-}\rho-\rho\sigma_{12}^{+}\sigma_{12}^{-})\nonumber\\
&{}+k_2(2\sigma_{02}^{-}\rho\sigma_{02}^{+}-\sigma_{02}^{+}\sigma_{02}^{-}\rho-\rho\sigma_{02}^{+}\sigma_{02}^{-}),
\end{alignat*}
where $\sigma_{12}^{-}$ and $\sigma_{12}^{+}$ are the same as above, and
\begin{equation} \sigma_{02}^{-}=|0\rangle\langle2|,\qquad
\sigma_{02}^{+}=|2\rangle\langle0|.
\end{equation}
By direct calculation, one can verify that the evolution of this
master equation can be expressed by using the following Kraus operators:
\begin{equation}
\rho(\tau)=\sum_{i=0}^{2}A_i\rho_0A_i^{\dag},
\end{equation}
where
\begin{alignat}{5}
A_0&{}={\rm diag}\{1,1,\sqrt{1-\gamma_1-\gamma_2}\},\nonumber\\
A_1&{}=\sqrt{\gamma_1}|0\rangle\langle2|,\nonumber\\
A_2&{}=\sqrt{\gamma_2}|1\rangle\langle2|,
\end{alignat}
and
\begin{alignat*}{5}
\gamma_1&{}=\frac{k_2}{k_1+k_2}\left[1-e^{-2(k_1+k_2)\tau}\right]&{}=2k_2\tau+O(\tau^2),\\
\gamma_2&{}=\frac{k_1}{k_1+k_2}\left[1-e^{-2(k_1+k_2)\tau}\right]&{}=2k_1\tau+O(\tau^2).
\end{alignat*}
Both $\gamma_1$ and $\gamma_2$ are of first order in $\tau$.

For the $V$-pattern, the Kraus expression has been found in
\cite{CW07}, which is given as
\begin{equation}
\rho(\tau)=\sum_{i=0}^{2}A_i\rho_0A_i^{\dag},
\end{equation}
where
\begin{alignat*}{5}
A_0&{}={\rm diag}\{1,\sqrt{1-\gamma_1},\sqrt{1-\gamma_2}\},\nonumber\\
A_1&{}=\sqrt{\gamma_1}|0\rangle\langle1|,\nonumber\\
A_2&{}=\sqrt{\gamma_2}|0\rangle\langle2|,
\end{alignat*}
and
\begin{alignat*}{5}
\gamma_1&{}=1-e^{-2k_1\tau}&{}=2k_1\tau+O(\tau^2),\\
\gamma_2&{}=1-e^{-2k_2\tau}&{}=2k_2\tau+O(\tau^2).
\end{alignat*}
Again, both $\gamma_1$ and $\gamma_2$ are of first order $\tau$.

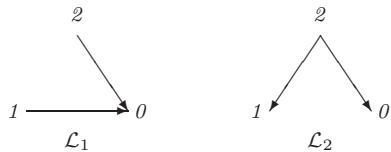
\begin{figure}[h!]
\vskip-\baselineskip
\centerline{\footnotesize\unitlength0.8\unitlength%
\begin{tabular}{cc}
\begin{picture}(100,80)(0,20)
\put(50,66){\makebox(0,0){$\mathit 2$}}
\put(20,20){\makebox(0,0){$\mathit 1$}}
\put(80,20){\makebox(0,0){$\mathit 0$}}
\put(26,20){\vector(1,0){48}}
\put(50,56){\vector(2,-3){24}}
\end{picture}
&
\begin{picture}(100,80)(0,20)
\put(50,66){\makebox(0,0){$\mathit 2$}}
\put(20,20){\makebox(0,0){$\mathit 1$}}
\put(80,20){\makebox(0,0){$\mathit 0$}}
\put(50,56){\vector(-2,-3){24}}
\put(50,56){\vector(2,-3){24}}
\end{picture}
\\
$ \cL_1$
&
$\cL_2$
\end{tabular}}
\caption{Classical channels for trits: the arrows indicate the allowed transitions.}
\label{fig:trits}
\end{figure}

We also introduce the classical counterparts of these two channels
which are shown in Fig. \ref{fig:trits}.  Here the arrows indicate the
allowed transitions.
The classical channels $\cL_1$ and $\cL_2$ correspond to
the amplitude damping channels $V$ and $\Lambda$, respectively.

For the $V$-channel and the $\Lambda$-channel, we construct codes
using similar ideas as shown above.  For single-error-correcting
codes, we can still use the idea of self-complementary codes based on
the corresponding classical codes, which have been studied in
\cite{Rob78} and \cite{Klo83}.  Unfortunately, none of these
constructions can be adapted to make self-complementary codes.  For
codes with short lengths, one can use numerical search.

For multi-error-correcting codes, we want to choose the basis of the
inner code so that one single damping error will project the state to
an orthogonal subspace.  Let $T_0$ and $T_1$ be the set of binary
strings of length $m$ with even and odd parity, respectively:
\begin{equation}
T_i = \{x_1 x_2 \ldots x_m \in \{0,1\}^m\mid x_1 \oplus x_2 \oplus \ldots \oplus x_m = i \}.
\end{equation}
For the channel $\cL_1$, a binary $0$ is mapped to the channel symbols
$\mathit 1$ and $\mathit 2$, and a binary $1$ is mapped to $\mathit
0$.  For the channel $\cL_2$, a binary $0$ is mapped to the channel
symbols $\mathit 0$ and $\mathit 1$, and a binary $1$ is mapped to
$\mathit 2$.  Then $T_0$ and $T_1$ are mapped to the set of codes
$S_0$ and $S_1$ which are two sets of possible inner codes.

\begin{example}
For $m=2$, we have
\begin{equation}
T_0 = \{00, 11\},\quad T_1 = \{01, 10\}.
\end{equation}
For the channel $\cL_1$, we get
\begin{equation}
S_0 = \{\mathit{00},\mathit{11},\mathit{12},\mathit{21},\mathit{22}\},
\quad S_1 = \{\mathit{01},\mathit{02},\mathit{10},\mathit{20}\},
\end{equation}
and for the channel $\cL_2$ channel:
\begin{equation}
S_0 = \{\mathit{00},\mathit{01},\mathit{10},\mathit{11},\mathit{22}\},
\quad S_1 = \{\mathit{01},\mathit{02},\mathit{10},\mathit{20}\}
\end{equation}
\end{example}

We have the following theorem.
\begin{theorem}
For any codeword $\mathbf{w}\in S_i$, $i = 0,1$, when an error of the
corresponding channel occurs, the resulting string $\mathbf{v}$ will be in
$S_{1-i}$.
\end{theorem}
\begin{IEEEproof}
We only prove the case when the channel is $\cL_1$, the proof for the
channel $\cL_2$ follows using the same argument.  Let $\mathbf{w} =
w_1 w_2 \ldots w_m$ and $\mathbf{v} = v_1 v_2 \ldots v_m$, and let the
error occur at position $t$. Then $w_t$ is either $\mathit{1}$ or
$\mathit{2}$, while $v_t =\mathit{0}$, and $w_k = v_k$, for all $k\ne
t$.  Suppose that in our construction, $w_1 \ldots w_m$ corresponds to
the binary string $a_1 \ldots a_m$, and $v_1 \ldots v_m$ corresponds
to the binary string $b_1 \ldots b_m$, then we have $a_t = 0$, $b_t =
1$, and $a_k = b_k$, for $k\ne t$. Since $\mathbf{a} \in T_i$, we know
that $\mathbf{b}\in T_{1-i}$, and hence $\mathbf{v}\in S_{1 - i}$.
\end{IEEEproof}
\begin{corollary}
For any codeword $\mathbf{w}\in S_i$, $i = 0,1$, when an odd number of
errors of the corresponding channel occur, the resulting string
$\mathbf{v}$ will be in $S_{1-i}$;  when an even number of errors of
the corresponding channel occur, the resulting string $\mathbf{v}$ will be
in $S_i$.
\end{corollary}

With arguments similar as those in the proof of Theorem \ref{th:mec},
for $i = 0, 1$, if we use the quantum code $\cQ_i$ spanned by
$\{\ket{\mathbf{u}}\colon \mathbf{u} \in S_i\}$ as the inner code and
an $[\![n,k,t+1]\!]_{|S_i|}$ code as the outer code, we get an
$[\![nm,K^k]\!]$ code that corrects $t$ errors for the corresponding
quantum AD channel.

Finally we will give an explicit expression for $|S_i|$ in terms of
$m$. To show its dependency on $m$, we write $\alpha_m = |S_0|$ and
$\beta_m = |S_1|$.
\begin{theorem}
\begin{equation}
\alpha_m = |S_0| = \frac{1}{2}(3^m + 1) \quad\text{and}\quad \beta_m = |S_1| = \frac{1}{2}(3^m - 1).
\end{equation}
\end{theorem}
\begin{IEEEproof}
For any string $\mathbf{a} = a_1 a_2 \ldots a_m \in T_0$, either $a_m
= 0$ or $a_m = 1$. For $a_m = 0$, the strings in $S_0$ that $a$ maps
to are just the strings $a_1 \ldots a_{m-1}$ maps to, concatenated
with a single symbol chosen from two options ($\mathit{1},\mathit{2}$
for $\cL_1$, and $\mathit{0},\mathit{1}$ for $\cL_2$).
For $a_m = 1$, the strings $\mathbf{a}$ maps to are those that $a_1
\ldots a_{m-1}$ maps to, concatenated with a fixed symbol. So we have
the recurrence relation
\begin{equation}
\label{eq:recurr1}
\alpha_m = 2\alpha_{m-1} + \beta_{m-1}.
\end{equation}
We also have
\begin{equation}
\label{eq:recurr2}
\alpha_m + \beta_m = 3^m.
\end{equation}
Solving Eqs.~\eqref{eq:recurr1} and \eqref{eq:recurr2} together with
the initial condition $\alpha_1 = 1$ and $\beta_1 = 2$, we have
\begin{equation}
\label{eq:size}
\alpha_m = \frac{1}{2}(3^m + 1)\quad\text{and}\quad \beta_m = \frac{1}{2}(3^m - 1),
\end{equation}
which proves the theorem.
\end{IEEEproof}
To achieve the maximal size of the constructed code, we should always
choose $\{\ket{\mathbf u}\colon \mathbf u \in S_0\}$ as the inner code.
\begin{example}
We take $m=2$, and the outer code is $[\![5,1,3]\!]_5$ (see, e.\,g.,
\cite{F97}), which is 
\begin{alignat}{5}
\ket{k} \mapsto \frac{1}{5\sqrt 5}\sum_{p,q,r=0}^4\omega^{k(p+q+r)+pr}\ket{p+q+k}\otimes\ket{p+r}
{}\otimes \ket{q+r}\otimes \ket{p}\otimes \ket{q},
\end{alignat}
where $\omega = \exp(2\pi i/5)$. We substitute
$\ket{0},\ket{1},\dots,\ket{4}$ with
$\ket{00},\ket{11},\ket{12},\ket{21}$, and $\ket{22}$, respectively,
and get a $(\!(10,5)\!)_3$ code which corrects $2$ errors of the
channel $V$. If we substitute $\ket{0},\ket{1},\dots,\ket{4}$ with
$\ket{00},\ket{01},\ket{10},\ket{11}$, and $\ket{22}$, we get a
$(\!(10,5)\!)_3$ $2$-code for the channel $\Lambda$.  In comparison,
the best stabilizer code of length $10$ which corrects $2$ errors is
$[\![10,1,5]\!]_3$.
\end{example}
Other examples for which our construction outperforms stabilizer codes
include the $(\!(12,25)\!)_3$ and $(\!(14,125)\!)_3$ $2$-codes constructed from
outer codes $[\![6,2,3]\!]_5$ and $[\![7,3,3]\!]_5$, while the best
corresponding stabilizer codes are $[\![12,2,5]\!]_3$ and
$[\![14,4,5]\!]_3$.

\section*{Acknowledgments}
Z. Wei is supported by the Singapore National Research Foundation
under NRF RF Award No. NRFNRFF2013-13.  Z.-Q. Yin is funded by the
NBRPC (973 Program) 2011CBA00300 (2011CBA00302), NNSFC 11105136,
61033001, and 61361136003.  B. Zeng is supported by NSERC and CIFAR.



%

\bibliographystyle{IEEEtranS}
\bibliography{QAD}

\end{document}